\documentclass[12pt]{article}
\usepackage{a4wide,amssymb,amsmath}
\usepackage{slashed}
\usepackage{wasysym}
\usepackage{enumerate}
\def\Fint{\rlap{$\Biggl\rfloor$}\Biggl\lceil}
\makeatletter
\newcommand{\xequal}[2][]{\ext@arrow 0055{\equalfill@}{#1}{#2}}
\def\equalfill@{\arrowfill@\Relbar\Relbar\Relbar}
\makeatother

\begin{document}

\begin{titlepage}

\begin{flushleft}
IZTECH-P/2012-02\\
\end{flushleft}

\begin{center}
\vspace{1cm}

{\Large \bf
Gravi-Natural Higgs and Conformal New Physics}

\vspace{0.5cm}

{\bf D. A. Demir}

\vspace{.8cm}

{\it Department of Physics, {\.I}zmir Institute of Technology, TR35430, {\.I}zmir, Turkey}

\vspace{.8cm}

(\today)

\end{center}
\vspace{1cm}

\begin{abstract}
\medskip
We study gauge hierarchy problem of the Standard Model ($SM$) not by introducing new physics at the electroweak scale but by utilizing gravitational frames, frames generated by conformal transformations, as a renormalization medium.  The essence of the naturalization
mechanism is that a particular conformal transformation with respect to the Higgs field completely decouples Higgs from matter fields so that the terms causing quadratic divergences become genuine benign. In this frame, we analyze stability, unitarity and renormalizability of the interactions,
and construct the effective $SM$ theory at the electroweak scale by taking into account the conformal anomalies. We scrutinize certain salient features of the framework, and discuss $h \rightarrow \gamma\gamma$ decay as an application.

The proposed mechanism naturalizes not only the $SM$ but also its extensions if they involve no new mass scales and exhibit strict conformal invariance. The new physics thus forms a conformal field theory (CFT), and can involve additional fermion generations, extra gauge groups as well as all kinds of $SM$ singlets. Certain phenomena, for instance, CP-violating Higgs sector, Dark Matter and Inflation, can be modeled with less free parameters. Scalar and vector singlets can directly interact with the $SM$ spectrum, and can be searched for in certain scattering processes such as the invisible decays of the Higgs boson.

\end{abstract}

\bigskip
\bigskip

\begin{flushleft}
IZTECH-P/2012-02
\end{flushleft}

\end{titlepage}

\tableofcontents

\section{Introduction}
\label{sec1}
The ATLAS and CMS experiments have discovered a Higgs boson candidate of mass
around $125\, {\rm GeV}$ \cite{higgs}. This completes
the Standard Model of strong and electroweak interactions ($SM$) to be a quantum
field theory describing Nature around the Fermi scale.  The $SM$, however, cannot
be a complete theory as it does not cover gravitation, as it does
not have a candidate for particle Dark Matter, as it does not include cosmic inflation, 
and as it does not do good for several other phenomena. More vitally and fundamentally than 
these, however, it suffers from the gauge hierarchy problem \cite{thooft1,veltman1},
that is,  the $SM$ is plagued by the perplexing naturalness problem
that the shortest-lived quantum fluctuations give the largest contributions to Higgs
boson mass \cite{wilson}. In the classical regime, the squared-mass parameter $m_{H}^2$ of the Higgs field,
being the only dimensionful parameter in the model, sets the Fermi scale. This scale,
however, gets destabilized towards larger and larger values when quantum fluctuations
of shorter and shorter durations come into play. This feature manifestly ensures that
the $SM$ Higgs sector needs stabilization against quantum fluctuations of the fields
it couples to. It is this stabilization problem, the gauge hierarchy problem,
which has been the deriving force behind various models of new physics constructed to
rehabilitate the $SM$ beyond Fermi energies. The ATLAS and CMS experiments are continuing
to search for their signals.

In the present work, we develop an alternative approach to naturalization of the $SM$ Higgs
sector. Our approach differs from the existing ones in utilizing gravitational frames instead
of introducing purposefully-structured new particles and symmetries at the electroweak scale.
The essence of the naturalization mechanism is that a particular conformal transformation with 
respect to the Higgs field \cite{einstein-jordan, r-dick, demir-onceki,demir} can modify its interactions with matter fields so that the terms
dangerous for naturalness can be avoided. This property proves crucial for achieving a natural effective
field theory at the electroweak scale. In developing this approach, we have excogitated  the viewpoint
that the Higgs field, being the generator of particle masses, should have a certain connectedness with
the gravitational dynamics more directly than the other fields in the $SM$ spectrum. We realize this
viewpoint by carrying Higgs field to different frames in which gravitational dynamics changes
with the conformal factor. In our approach, conformal factor is identified with the norm
of the Higgs field. (For some of the interesting approaches in the literature see
\cite{wetterich} and \cite{otherx}. The gravitational asymptotic safety arguments of \cite{wetterich}
predicts Higgs mass to be in the ballpark of the LHC result \cite{higgs}.)

The paper is organized as follows. In Sec. \ref{sec2} below, we briefly review the Higgs naturalness problem
by explicitly determining the sensitivities of the model parameters to the ultraviolet (UV) boundary.
For use in subsequent sections, we interpret the $SM$ interactions in curved geometry of a strictly
classical metric field. We compute effects of the matter loops on the gravitational and material
sectors to determine the effective $SM$ theory at the electroweak scale. At the end of the section,
by a detailed discussion of the renormalization properties of the model parameters, we restate
the Higgs naturalness problem.

In Sec. \ref{sec3}, we propose a new framework for solving the Higgs naturalness problem. The idea is to use
the gravitational frames, the frames that are related through conformal transformations, as a
regularization medium. In fact, we show that the frame which decouples Higgs field from rest
of the matter fields \cite{demir} naturally avoids the terms dangerous for naturalness, and leads
to naturalization of the Higgs sector. The naturalness here arises from the global scaling invariance
induced by the conformal transformations. We provide a detailed analysis of unitarity and
renormalization properties of the $SM$. The conformal anomalies break the scaling symmetry
softly, that is, by inducing new interactions right at the electroweak scale. We construct the
effective $SM$ theory at the electroweak scale by determining effective potential and effective
vertices, separately. We then analyze $h \rightarrow \gamma \gamma$ decay to illustrate the renormalization programme.

In Sec. \ref{sec3} again, we show that the ghosty nature of
the Higgs field in the Gravic frame enforces new physics, if any, to be a CFT. We then
explore what forms of new physics can add to the $SM$ interactions, and find that sequential
generations, new gauge groups as well as generic singlet sectors are admissible. Accordingly, we 
examine a number of phenomena, massive neutrinos, CP violation, Dark Matter and Inflation, and
show that they are all facilitated and their descriptions involve less free parameters. Specifically,
we find that scalar and (gauge or non-gauge) vector singlets can couple to the $SM$ via
Higgs, hypercharge and fermion portals, and they can lead to viable models of Dark Matter 
and Inflation, and can be searched for in the invisible width of the Higgs boson \cite{higgs}.

In Sec. \ref{sec4} we conclude.

\section{Higgs Naturalness Problem}
\label{sec2}
We focus on the $SM$ with its known particle spectrum and known interaction
schemes \cite{sm}. We do not consider extended models involving new scales (higher than the electroweak scale). Because
of the relevance of conformal transformations, we include gravitational interactions by interpreting
the $SM$ dynamics as taking place in the curved geometry of the metric tensor $g_{\mu\nu}$.
The geometric sector is described by the usual Einstein-Hilbert term. This setup we name {\it Standard frame}
to mean that gravitational sector is in the Einstein frame, and the material sector is the $SM$ (though, as will
be shown later, its certain extensions are allowed).

The Higgs field, the only scalar field in the $SM$, can be parametrized  as
\begin{eqnarray}
\label{higgs}
H = \frac{\Phi}{\sqrt{2}} {\bf U}\left(\vec{\varphi}\right) \left(\begin{array}{c} 0 \\ 1\end{array}\right)
\end{eqnarray}
where ${\bf U}\left(\vec{\varphi}\right)\in SU(2)$ encodes the three Goldstone bosons $\varphi_{1,2,3}$
swallowed by the electroweak gauge bosons in acquiring their masses.

The $SM$ plus gravity setup is highly minimal in that it contains only two fundamental scales, which are conveniently collected
in the energy density
\begin{eqnarray}
\label{tree-pot-ST}
V(\Phi,g) = V_0 +  \frac{1}{2} m_{H}^2 \Phi^2 + \frac{1}{4} \lambda_{H} \Phi^4 - \frac{1}{2} M_{Pl}^2 {{R}}(g)
\end{eqnarray}
where $V_0$ is a primordial (uncalculable) constant energy density. The squared-mass parameter of the Higgs
field, $m_H^2$, sets the characteristic scale of the matter sector via the Fermi constant $G_{F}= \lambda_{H}/\left(\sqrt{2} |m_H^2|\right)$.
The Planck mass, $M_{Pl}$, sets the gravitational scale via the Newton constant $G_N = 1/(8\pi M_{Pl}^2)$. These two scales, $G_F$ and $G_N$, constitute, respectively, electroweak and ultraviolet ends of the total energy span.

The relation of $G_F$ to $m_H^2$, given above, follows from minimization of $V(\Phi,g)$ with respect to $\Phi$. In fact,
$G_F = 1/(\sqrt{2} \langle \Phi^2 \rangle)$ where $V(\Phi,g)$ is minimized at $\langle \Phi^2 \rangle = - m_H^2/\lambda_H$
for $m_H^2 < 0$. Phenomenologically, $m_{H}^2/\lambda_H$ is fixed in terms of the known value of $G_F$. This is an intimately
classical analysis. For the $SM$ to make sense in regard to weak interactions, $m_{H}^2$ must stay put at the scale $G_F$ even
after the inclusion of quantum effects.

In quantum theory, the Higgs VEV  $\langle \Phi \rangle$ could still be determined by a similar minimization procedure if
$V(\Phi,g)$ is improved to incorporate quantum fluctuations of fields. This improved potential, the effective potential \cite{coleman-weinberg},
adjoins quantum fluctuations into $V(\Phi,g)$ such that its deepest minimum and particles masses generated therein
properly encode the quantum effects. We shall compute the effective potential below.

\subsection{Physical UV Cutoff}
\label{sec21}
Our goal is to construct the $SM$ effective field theory at the scale $\Lambda_{\texttt{EW}} \gtrsim m_H$. This is
the renormalization scale. The UV scale, denoted by $\Lambda_{\texttt{UV}}$, is the ultimate validity limit of the $SM$, and it
can be as high as the Planck scale $M_{Pl}$ if the $SM$ proves to be the correct model of Nature from $G_F$ all the way down to $G_N$.
Considering the $SM$ (or its extensions not involving any new mass scales), the UV scale becomes the Planck scale itself, $\Lambda_{\texttt{UV}}\sim M_{Pl}$, and hence, $\Lambda_{\texttt{UV}}$ can be regarded as a {\it physical scale} rather than a formal momentum cutoff on the loop integrals. We reiterate that $\Lambda_{\texttt{UV}}$ is as physical as any other parameter in the theory, and the effective theory at the electroweak scale
proves natural if its parameters do not escape to the scale of $\Lambda_{\texttt{UV}}$ under quantum corrections. The perturbation
expansion proceeds with loop integrals extending from $\Lambda_{\texttt{EW}} \gtrsim m_H$ up to $\Lambda_{\texttt{UV}}\sim M_{Pl}$, and
their $\Lambda_{\texttt{EW}}$ and $\Lambda_{\texttt{UV}}$ dependencies are physical relations.

In computing the quantum corrections, we analyze fields in two distinct classes:
\begin{itemize}
\item Classical fields: We take metric tensor $g_{\mu\nu}$ to be strictly classical;
it possesses no quantum fluctuations for all energy scales up to $\Lambda_{\texttt{UV}}\sim M_{Pl}$. Its
quantum nature becomes significant for energies above $M_{Pl}$, in the framework of string theory.

\item Quantal fields: We take matter fields to be all quantum fields. They are conveniently decomposed into
slow and fast components
\begin{eqnarray}
\label{fluctuations}
\Phi = \phi + \delta\phi\,,\; \Psi = \psi + \delta\psi\,,\; V_{\mu} = \upsilon_{\mu} + \delta\upsilon_{\mu}
\end{eqnarray}
where $\phi$, $\psi$, $\upsilon_{\mu}$ are long-wavelength fields possessing energies at most $\Lambda_{\texttt{EW}} \gtrsim m_H$ whereas
$\delta\phi$, $\delta\psi$, $\delta\upsilon_{\mu}$ are high-frequency fields extending in energy from $\Lambda_{\texttt{EW}}$  up to $\Lambda_{\texttt{UV}}\sim M_{Pl}$. These are the quantum fluctuations about the slow fields $\phi$, $\psi$, $\upsilon_{\mu}$ such that the
latter satisfy the classical equations of motion modulo the quantum corrections.
\end{itemize}

\subsection{Effective Potential}
\label{sec22}
The effective action at the electroweak scale is obtained by integrating out the quantum fluctuations \cite{duff-christensen}. The generating functional of field correlators,  which involves fields of all frequencies, undergoes the reduction process
\begin{eqnarray}
\label{part-func-ST}
{Z} &=& \Fint \left[{\mathfrak{D}} \Phi\,  {\mathfrak{D}} \overline{{\Psi}}\, {\mathfrak{D}} {{\Psi}}\, {\mathfrak{D}} V_{\mu}\right]_{{g}} e^{{i} {{S}}\left[\Phi, {\Psi}, V, {g}\right]}\nonumber\\
&\xequal{{\text{Eq.}(\ref{fluctuations})}}& \Fint \left[{\mathfrak{D}} \phi\,  {\mathfrak{D}} \overline{{\psi}}\, {\mathfrak{D}} {{\psi}} {\mathfrak{D}} \upsilon_{\mu}\right]_{{g}}
\left[{\mathfrak{D}}\left(\delta \phi\right)  {\mathfrak{D}}(\overline{\delta {\psi}}) {\mathfrak{D}}({\delta {\psi}}) {\mathfrak{D}}\left(\delta \upsilon_{\mu}\right)\right]_{{g}} e^{{i} {{S}}\left[\Phi, {\Psi}, V, {g}\right]}\nonumber\\
&=& \Fint \left[{\mathfrak{D}} \phi\,  {\mathfrak{D}} \overline{{\psi}}\, {\mathfrak{D}} {{\psi}}\, {\mathfrak{D}} \upsilon_{\mu}\right]_{{g}} e^{{i} {{S}}_{eff}\left[\phi, {\psi}, \upsilon, {g}; \Lambda_{\texttt{EW}}\right]}
\end{eqnarray}
to involve only the long-wavelength fields $\phi$, $\psi$, $\upsilon_{\mu}$ through ${{S}}_{eff}\left[\phi, {\psi}, \upsilon, {g}; \Lambda_{\texttt{EW}}\right]$. The effective action encodes the $SM$ interactions with all parameters renormalized at the
electroweak scale $\Lambda_{\texttt{EW}}$. The $SM$ theory thus formed contains, not the tree-level potential (\ref{tree-pot-ST}),
but the effective potential
\begin{eqnarray}
\label{1loop-pot-ST}
V_{eff}(\phi,g; \Lambda_{\texttt{EW}}) = \overline{V_0} +  \frac{1}{2} \overline{m_{H}^2} \phi^2 + \frac{1}{4} \overline{\lambda_{H}} \phi^4 - \frac{1}{2} \left(\overline{M_{Pl}^2} - \overline{\lambda_R} \phi^2\right) {{R}}(g)
\end{eqnarray}
which holds for small curvatures (quadratic and higher order terms in curvature tensors are neglected) at low energies (it involves long-wavelength fields $\phi$ and $g_{\mu\nu}$). The renormalized masses and couplings in it can be determined by considering generic Higgs interactions of the form
\begin{eqnarray}
\label{generic-couplings}
- \frac{1}{2} \lambda_V \Phi^2 g^{\mu \nu} V_{\mu} V_{\nu} - \frac{1}{\sqrt{2}} \lambda_{\Psi} \Phi \overline{\Psi} \Psi
\end{eqnarray}
which correspond to the $SM$ interactions in the unitary gauge. For these couplings, with one loop accuracy, the renormalized parameters in (\ref{1loop-pot-ST}) read as
\begin{eqnarray}
\label{renormed_Mpl-ST}
{{\overline{M_{Pl}^2}}} &=& {M_{Pl}}^2 + \frac{1}{(4 \pi)^2}\, \left[ \left( \frac{1}{6} - \frac{n_1^0}{6} - \frac{n_1}{4} +  \frac{n_{1/2}}{3} \right)\Lambda_{\texttt{UV}}^2 + \frac{1}{6} m_{H}^2 \log \frac{\Lambda_{\texttt{EW}}^2}{\Lambda_{\texttt{UV}}^2 }\right],\\
\label{renormed_Mphi-ST}
{{\overline{m_{H}^2}}} &=& m_{H}^2 + \frac{1}{(4 \pi)^2}\, \left[ \left( 3 \lambda_{H} + 3 \lambda_V - 2 \lambda_{\psi}^2 \right) \Lambda_{\texttt{UV}}^2
+ 3 \lambda_{H} m_{H}^2 \log \frac{\Lambda_{\texttt{EW}}^2}{\Lambda_{\texttt{UV}}^2} \right],\\
\label{renormed_V0-ST}
{{\overline{V_0}}} &=& V_0 + \frac{1}{(4 \pi)^2}\, \left[ \frac{1}{4} (n_F - n_B) \Lambda_{\texttt{UV}}^4  + \frac{1}{2} m_{H}^2 \Lambda_{\texttt{UV}}^2 + \frac{1}{4} {(m_{H}^2)}^2 \log \frac{\Lambda_{\texttt{EW}}^2}{\Lambda_{\texttt{UV}}^2 }\right],
\end{eqnarray}
and
\begin{eqnarray}
\label{renormed_zeta-ST}
{{\overline{\lambda_R}}} &=& \frac{1}{(4 \pi)^2} \left( - \frac{1}{2} \lambda_{H} + \frac{1}{4} \lambda_{V} - \frac{1}{6} \lambda_{\psi}^2 \right)
\log \frac{\Lambda_{\texttt{EW}}^2}{\Lambda_{\texttt{UV}}^2 },\\
\label{renormed-lamphi-ST}
\overline{\lambda_{H}} &=& \lambda_{H} + \frac{1}{(4 \pi)^2}\left( 9 \lambda_{H}^2 + 3 \lambda_V^2 - \lambda_{\psi}^4\right) \log \frac{\Lambda_{\texttt{EW}}^2}{\Lambda_{\texttt{UV}}^2 }\,.
\end{eqnarray}
In these expressions, the integers $n_B$, $n_F$, $n_1^0$, $n_1$, $n_{1/2}$ count, respectively, the total numbers of bosons, fermions,
massless vectors, massive vectors and Dirac fermions. Their numerical values can be determined from the $SM$ spectrum.

The way the model parameters renormalize in (\ref{renormed_Mpl-ST}-\ref{renormed-lamphi-ST}) reveal a number of important aspects
of the $SM$ in the UV:
\begin{itemize}
\item As revealed by (\ref{renormed_Mpl-ST}), $M_{Pl}$ receives at most a few $\%$ correction even for $ \Lambda_{\texttt{UV}} \sim M_{Pl}$. This means that the scale of quantum gravity does not change with quantum fluctuations of matter fields. In other words, the scale at which the quantum fluctuations of metric tensor dominate is radiatively stable.

\item As is seen from (\ref{renormed_Mphi-ST}), quantum corrections to Higgs mass-squared vary quadratically and
logarithmically with the UV scale. Quadratic sensitivity to $\Lambda_{\texttt{UV}}$ is a complete disaster because, with ${{\overline{m_{H}^2}}} \sim \Lambda_{\texttt{UV}}^2$, the $SM$ completely fails to account for weak interactions. Essentially, quantum effects lift up the electroweak scale $\Lambda_{\texttt{EW}} \sim m_H$ near the UV scale $\Lambda_{\texttt{UV}}$, and hence, the $SM$ gets entirely displaced from the Fermi scale. This quantum instability is a characteristic feature of scalar field theories \cite{wilson}, and cannot be altered by changing the regularization scheme. (Quadratic divergence becomes a pole in $d=2$ dimensions within dimensional regularization \cite{bardin-passarino}. Likewise, quartic divergence becomes a pole in $d=0$, and logarithmic divergence becomes a pole in $d=4$ dimensions.)

\item Similar to ${{\overline{m_{H}^2}}}$, the renormalized vacuum energy $\overline{V_0}$ also exhibits a strong sensitivity to the UV boundary. As indicated by (\ref{renormed_V0-ST}), it grows quartically, quadratically and logarithmically with $\Lambda_{\texttt{UV}}$, and hence, it gets completely destabilized from its classical value $V_0$ (even if this value agrees with observations \cite{ccp-exp}). This is yet another source of unnaturalness induced by quantum effects; it is the cosmological constant problem \cite{ccp-weinberg}.

\item The dimensionless couplings are renormalized logarithmically. The quantum corrections induce a non-minimal coupling term $\overline{\lambda_R}$ (which could have been included from the beginning by adding $\frac{1}{2} \lambda_R \Phi^2 R(g)$ to $V(\Phi,g)$). The Higgs quartic coupling, $\overline{\lambda_{H}}$, also renormalizes logarithmically.

\end{itemize}
The strong UV sensitivities of the dimensionful parameters render the $SM$ completely unnatural \cite{nima,peskin}. For achieving naturalness, the theory must be completed by some `new physics' above the electroweak scale. The naturalness-driven `new physics' candidates are the models based on extra dimensions, supersymmetry and technicolor.

\section{Higgs Naturalness Frame}
\label{sec3}
Having explicated the problem, we ask: In the setup of the $SM$ plus gravity, can one excogitate a way of taming the quadratic divergences? The answer is affirmative. Indeed, we shall show that renormalization properties of the model parameters suggest yet another approach to the naturalness problem, distinct from the known ones.

For this purpose, we focus on ${{\overline{m_{H}^2}}}$ given in (\ref{renormed_Mphi-ST}). The quadratic divergences here (the terms proportional to $\Lambda_{\texttt{UV}}^2$) involve the Higgs coupling to itself ($\lambda_{H}$), Higgs coupling to vector bosons ($\lambda_V$), and Higgs coupling to fermions ($\lambda_{\Psi}$). It is through these couplings of the Higgs field that the loops of $\delta\phi$, $\delta\psi$ and $\delta\upsilon_{\mu}$ give rise to quadratic divergences. Therefore, a judicious modification in these interaction terms could modify structures of the divergences. Actually, one is to determine if there is any physically consistent modification of these Higgs interactions so that quadratic divergences are alleviated. Essentially, we think of a transformation on the fields in the $SM$ action such that the Higgs field is erased from those terms dangerous for naturalness. Hence, the requisite transformation must be something like
\begin{eqnarray}
\label{trans-expect}
\lambda_{H} \Phi^4 \rightarrow \lambda_{H} M^4\;,\;\; \lambda_V \Phi^2 g^{\mu\nu}V_{\mu} V_{\nu} \rightarrow \lambda_V M^2 g^{\mu\nu} V_{\mu} V_{\nu}\;,\;\; \lambda_{\Psi} \Phi \overline{\Psi} \Psi \rightarrow \lambda_{\Psi} M \overline{\Psi} \Psi
\end{eqnarray}
where we did not indicate possible transformations of the fields themselves. Here $M$ is a mass scale introduced to replace $\Phi$. Under (\ref{trans-expect}), the interactions in (\ref{tree-pot-ST}) and (\ref{generic-couplings}) become free of the Higgs field, and hence, the quantum fluctuations in (\ref{fluctuations}) cannot induce a potential for $\Phi$.

\subsection{Gravic Frame}
\label{sec31}
A careful look at the transformations in (\ref{trans-expect}) immediately reveals that they can actually be realized via conformal transformations of the metric and matter fields. Indeed, for implementing the desired changes in (\ref{trans-expect}), it suffices to apply the conformal transformations \cite{conf-trans}
\begin{eqnarray}
\label{conf-trans}
g_{\mu \nu} = \left(\frac{\Phi}{M_c}\right)^{-2} \tilde{g}_{\mu\nu}\,,\; \Psi =  \left(\frac{\Phi}{M_c}\right)^{3/2} \tilde{\Psi}
\end{eqnarray}
whose conformal factor is given by the Higgs field itself \cite{einstein-jordan,r-dick,demir}. The scale $M_c$ here arises for dimensionality reasons. To see that these transformations do indeed yield the desired changes in (\ref{trans-expect}), it suffices to examine the Higgs interactions \begin{eqnarray}
\label{changes}
&&\sqrt{-g} H^{\dagger} T^{i} T^{j} H \, g^{\mu \nu} V^{i}_{\mu} V^{j}_{\nu} \rightarrow \frac{1}{2} M_c^2 \sqrt{-\tilde{g}} \left( {\bf U}^{\dagger} T^{i} T^{j} {\bf U}\right)_{22}\, \tilde{g}^{\mu \nu} V^{i}_{\mu} V^{j}_{\nu}\nonumber\\
&&\sqrt{-g} \overline{\Psi_R} H^{\dagger} \left(\begin{array}{l} \Psi_L^{\prime}\\ \Psi_L\end{array}\right) \rightarrow  \frac{1}{\sqrt{2}} M_c \sqrt{- \tilde{g}} \left[ \left({\bf U}^{\dagger}\right)_{2 1} \overline{\tilde{\Psi}_R}\, \tilde{\Psi}_L^{\prime} + \left({\bf U}^{\dagger}\right)_{2 2} \overline{\tilde{\Psi}_R}\, \tilde{\Psi}_L\right]
\end{eqnarray}
which directly reduce to those in (\ref{trans-expect}) for $M=M_c$ in the unitary gauge ${\bf U} = {\bf 1}$. Under (\ref{conf-trans}), the Higgs-self interactions are also modified
\begin{eqnarray}
\label{changes2}
\sqrt{-g} \left[ -\frac{1}{2} g^{\mu\nu} \partial_{\mu}\Phi \partial_{\nu} \Phi - V(\Phi,g)\right]
\rightarrow \sqrt{-\tilde{g}} \left[ \frac{1}{2} K(\Phi) \tilde{g}^{\mu\nu} \partial_{\mu}\Phi \partial_{\nu} \Phi - {\tilde{V}}(\Phi,\tilde{g})\right]
\end{eqnarray}
so that the Higgs field develops a non-minimal kinetic function
\begin{eqnarray}
\label{kin-func}
K(\Phi) &=& M_{c}^2 \Phi^{-2} \left[ 6 M_{Pl}^2 \Phi^{-2} - 1\right]
\end{eqnarray}
along with a modified potential term
\begin{eqnarray}
\label{tree-pot-GR}
{\tilde{V}}(\Phi,\tilde{g}) &=& \frac{1}{4} \lambda_{H}  M_{c}^4 +  \frac{1}{2} m_{H}^2 M_{c}^4 \Phi^{-2} +  V_0 M_{c}^4 \Phi^{-4}
- \frac{1}{2} M_{Pl}^2 M_{c}^2 \Phi^{-2} \tilde{R}(\tilde{g})\,.
\end{eqnarray}
The conformal transformations (\ref{conf-trans}) give rise to striking modifications in the Higgs field's dynamics. A detailed analysis of the modifications, at the classical level, is given in the recent paper \cite{demir} (see also \cite{einstein-jordan,r-dick,demir-onceki}). Here
we discuss a few of them, for completeness. First, one notices that the conformal transformations (\ref{conf-trans}) do merely peal off $\Phi$ from matter vertices; they do not touch the gauge symmetry at all since the Goldstone matrix ${\bf U}(\vec{\varphi})$ is left behind to ensure gauge invariance. (In the calculations below, however, we find it convenient to work in the unitary gauge, ${\bf U} = {\bf 1}$.)

Next, one notices the striking changes in Higgs interactions. After the conformal transformations (\ref{conf-trans}), Higgs field gets completely decoupled from matter fields \cite{demir}. Indeed, as is seen from (\ref{changes}) and (\ref{changes2}), $\Phi$ does not couple to any of the vector bosons, leptons and quarks. Stating generally, after the (\ref{conf-trans}), the Higgs field $\Phi$ gets decoupled from all the fields whose masses are generated by the Higgs mechanism, and does not develop couplings to fields in other sectors due to their conformal invariance. Needless to say, the Goldstone bosons in ${\bf U}\left(\vec{\varphi}\right)$ continue to interact with the matter fields in the way required by gauge invariance. Apart from this Higgs-matter decoupling, the conformal transformations (\ref{conf-trans}) make Higgs to possess only gravitational interactions. It is therefore convenient to name the frame generated by (\ref{conf-trans}) as the {\it Gravic frame} to mean that gravitational sector is in the Jordan frame, $\Phi$ possesses only gravitational interactions, and vector bosons and fermions acquire hard masses with no breaking of gauge invariance \cite{demir}. This frame accommodates various features not found in Standard frame. To this end, one notes changes in the meanings of various interactions. Indeed, a comparative look at (\ref{tree-pot-ST}) and (\ref{tree-pot-GR}) reveals that certain energy components change their roles as one switches from Standard frame to Gravic frame. For instance, the vacuum energy in Standard frame transforms into quartic coupling in Gravic frame. Similarly, quartic coupling in Standard frame becomes vacuum energy in Gravic frame. These changes in the roles will be seen to critically affect the structures of the quantum corrections, in agreement with (\ref{trans-expect}).

\subsubsection{Stability}
\label{sec311}
The Higgs field $\Phi$ is a real scalar field in the Standard frame. However, as is seen from (\ref{changes2}), it becomes a ghost field in the Gravic frame. This is because $K(\Phi) > 0$ as $M_{Pl}^2 \gg \Phi^2$ for all relevant field configurations. The other fields in the $SM$ spectrum continue to have their usual particle characteristics. Gaining ghosty nature makes Higgs field a geometrical field \cite{geometrical1,geometrical2,geometrical3} tied up to the curvature of spacetime. It possesses negative kinetic energy, and hence, negative residue in its propagator, and hence, negative probability to be produced from the interacting vacuum. The probability can be made positive by reversing the sign of the particle energy. Then, however, Higgs field decreases its energy unboundedly by emitting normal, positive-energy particles ({\it e. g.} fermions and vector bosons in the $SM$). Thus, the vacuum state decays till it possesses infinite negative energy. This instability is a disaster. The way out from it, obviously, is to prohibit Higgs boson from coupling to normal matter \cite{vilenkin}. This is naturally accomplished in the $SM$ as a fundamental property of the Gravic frame. Indeed, in the Gravic frame, the $SM$ Higgs field is completely decoupled from leptons, quarks and vector bosons; it interacts only with the classical gravity.

The vacuum stability puts strong restrictions on possible new fields beyond the $SM$. These fields, if any, must follow the pattern in the $SM$ in their interactions with the Higgs field. In other words, they must exhibit strict conformal invariance in Standard frame. In particular, none of them must have a hard mass or any other dimensionful coupling because then they develop a direct coupling to the Higgs field through such dimensionful couplings after the conformal transformation (\ref{conf-trans}). Consequently, in the Gravic frame, for the ghosty Higgs to be admissible it is necessary that possible new fields beyond the $SM$ spectrum form a conformal-invariant sector. In other words, new physics, if any, must be a CFT. Its fields must acquire their masses and other dimensionful couplings from their conformal-invariant couplings to the Higgs field. For example, a scalar field $S$ cannot have a hard mass term; its mass is generated after electroweak breaking via the conformal coupling $S^2 H^{\dagger} H$ in the Standard frame. We shall discuss fields beyond the $SM$ in Sec. \ref{sec33}.

\subsubsection{Unitarity}
\label{sec312}
As explicated in (\ref{changes}), in the Gravic frame, fermions and vector bosons completely decouple from the Higgs field, and acquire, respectively, the hard masses
\begin{eqnarray}
M_{\Psi} = \frac{\lambda_{\Psi}}{\sqrt{2}} M_c\,,\; M_V = \sqrt{\lambda_V} M_c\,.
\end{eqnarray}
These changes in interactions severely affect the UV behavior. Indeed, one immediately notices that the very absence of the Higgs field in the vector boson sector gives rise to the unitarity problem. The reason is that,  the longitudinally-polarized vector bosons, corresponding to the Goldstone bosons in ${\bf U}(\vec{\varphi})$, become strongly-coupled at a scale not too far from $M_V$. In fact, the cross section $\sigma\left(V^{i}_{L} V^{j}_{L} \rightarrow V^{i}_{L} V^{j}_{L}\right)$ violates the unitarity limit for center-of-mass energies above $\sqrt{\frac{4 \pi}{\lambda_V}} M_V$ \cite{unitarity}. In Standard frame, Higgs boson restores the unitarity. In Gravic frame this is simply not possible. Interestingly, however, in Gravic frame, vector boson masses are proportional to the scale $M_c$ which has nothing to do with the electroweak scale. Thus, one can exploit this freedom to preserve the unitarity up to high energies by taking $M_c$ large enough. Consequently, one naturally imposes the bound
\begin{eqnarray}
\label{McMpl}
M_{c} \gtrsim M_{Pl}
\end{eqnarray}
if the SM is to be valid up to the Planck scale. Under this bound, $M_c$ nestles in the string territory, and hence, momenta of vector bosons cannot exceed their masses $M_V$ to cross the unitarity border.

Having $M_c$ inside the string territory, question arises as to whether quantum gravitational effects can render low-energy predictions futile.
In general, sizes of the threshold contributions vary with the sensitivity of the low-energy theory to the UV domain. In the present formulation, energies and momenta of particles, as they are described by a quantum field theoretical framework, cannot exceed $\Lambda_{\texttt{UV}} \sim M_{Pl}$, and hence, threshold effects are not expected to be sizeable. The reason for this is this. In the Gravic frame, the ultra-large terms in the action involve $M_c^4$ or $\Lambda_{\texttt{UV}}^4$. However, at the end of the analysis, after performing inverse of the conformal transformations (\ref{conf-trans}) (see the equation (\ref{conf-trans-back}) below) these large effects will be washed out by $1/M_c^{4}$ factors. Therefore, threshold effects, though they deceptively seem so, cannot be too large to spoil predictions in the low-energy theory.

\subsubsection{Renormalizability}
\label{sec313}
Massive non-Abelian vectors give rise to difficulties in the UV. Unitarity discussed above is just one of them. The other aspect concerns renormalizability.  Indeed, massive vectors render theory non-renormalizable. Unless their masses are induced by Higgs mechanism, which are not in the Gravic frame, they unavoidably lead to a non-renormalizable theory. In Gravic frame, the Higgs field is decoupled from the rest, and does not spoil renormalizability excepting the higher dimensional terms coming from its self-interaction potential. On the other hand, spinors and vectors, both being endowed with hard masses,  are not renormalizable \cite{renorm,renorm1}. This is revealed by the asymptotic behaviors of their propagators. Indeed, for large momenta, propagator of the Higgs scalar scales as $1/p^2$. For fermions and vectors there is no large momentum regime, and their propagators
\begin{eqnarray}
\label{propagate}
\frac{1}{\slashed{p} + M_{\Psi}} \rightarrow \frac{1}{M_{\Psi}} - \frac{\slashed{p}}{M_{\Psi}^2}\;,\;\;
\frac{1}{p^2 + M_V^2} \left( g_{\mu\nu} + \frac{p_{\mu} p_{\nu}}{M_V^2}\right) \rightarrow  \frac{g_{\mu\nu}}{M_V^2} + \frac{1}{M_V^4} \left(p_{\mu} p_{\nu} - p^2 g_{\mu\nu} \right)
\end{eqnarray}
are given by their Compton wavelengths even for momenta close to the Planck scale. (If large momentum  regime were possible, these propagators would scale as $1/\slashed{p}$ and ${p_{\mu} p_{\nu}}/{M_V^2 p^2}$. The massive vector propagator would continue to spoil renormalizability.) These propagation properties follow from the unitarity bound (\ref{McMpl}). Indeed, being ultra-massive particles, these spinors and vectors cannot be made to propagate and scatter by supplying sub-Planckian energies relevant for quantum field theory. In this sense, for energies up to the Planck scale there arises no problem with unitarity. At the Fermi energies, in Gravic frame, these particles are essentially frozen, exhibit no dynamics, and  do not see the Higgs field.

The S-matrix elements involve loops of particles. The loops of spinors and vectors can have loop momenta as high as $\Lambda_{\texttt{UV}}$, and they may give significant contributions to various quantities such as vacuum energy, vector boson self-energy, fermion self-energy and vector boson-fermion-fermion vertices. All such loop amplitudes are divergent; they continue to be divergent even if certain renormalization conditions are imposed \cite{renorm1}. The reason is that, as suggested by the propagators in (\ref{propagate}), these loop amplitudes necessarily involve positive powers of $\Lambda_{\texttt{UV}}$, and hence, they render the theory non-renormalizable. For instance, at low energies, the renormalized vector boson mass is given by
\begin{eqnarray}
\label{amss}
\overline{M_V^2} = M_V^2 \left( 1 + \sum_{L} c_L \left(\frac{\alpha}{4\pi}\right)^{L} \left(\frac{\Lambda_{\texttt{UV}}}{M_V}\right)^{6L}\right)
\end{eqnarray}
where $L$ counts the number of loops in a given diagram. This expression is obtained by considering purely gluonic diagrams and neglecting the subleading momentum-dependent terms in the vector propagator (\ref{propagate}). With exact  propagator, the coefficients $c_L$,
which are expected to be ${\mathcal{O}}(1)$, would develop a mild dependence on  $\Lambda_{\texttt{UV}}$. The fermion loop contribution, which
goes like $\left(\Lambda_{\texttt{UV}}^2/(M_V M_{\Psi})\right)^{2L}$, exhibits a softer dependence on $\Lambda_{\texttt{UV}}$. The vector boson mass depends strongly on the UV scale. This is valid for other quantities, as well. This, however, does not need to cause an impasse. There are two reasons for this:
\begin{itemize}
\item As already discussed in Sec. \ref{sec21}, in the $SM$ plus gravity setup, the UV scale is a physical scale: $\Lambda_{\texttt{UV}} \simeq M_{Pl}$. It is nothing but the scale of strong gravity above which quantum field theory does not apply. The renormalized quantities (the
    vector boson mass and all the others) depend explicitly on $\Lambda_{\texttt{UV}}$ as they depend on any other parameter of the model.

\item If the $\Lambda_{\texttt{UV}}$ dependence destabilizes a parameter form its physically-expected scale then there arises  the naturalness problem. However, as guaranteed by the bound (\ref{McMpl}), the ratio $\Lambda_{\texttt{UV}}/M_c$ is small. This means that the all-loop sum in (\ref{amss}) cannot be too large to cause a significant renormalization of $M_V^2$. In other words, $\overline{M_V^2}$ is expected to be close to $M_V^2$.   (These points will be further discussed in Sec. \ref{sec322}.)
\end{itemize}
In consequence, in the $SM$ plus gravity setup of (\ref{tree-pot-ST}),  $\Lambda_{\texttt{UV}}$ is a physical scale, and
the renormalized parameters depend on it as they do on any other model parameter. The renormalized parameters make physical sense. In Standard frame, for instance,
renormalized parameters in (\ref{renormed_Mpl-ST})-(\ref{renormed-lamphi-ST}) do explicitly depend on $\Lambda_{\texttt{UV}}$
as a physical parameter. These dependencies are not natural, however. The reason is that they destabilize the electroweak scale.
Actually, what Gravic frame will do is to stabilize the model parameters against violent $\Lambda_{\texttt{UV}}$
dependencies by utilizing the scale $M_c$ through the conformal transformations. The renormalized vector mass in (\ref{amss}) provides an early example of this.

\subsection{Effective Action}
\label{sec32}
Having shown that the Gravic frame can remedy the naturalness problems arising in the Standard frame, we now construct the $SM$ effective field theory at the electroweak scale within the Gravic frame. We call this model $\widetilde{SM}$ to differentiate it from the $SM$ in the Standard frame. As in (\ref{part-func-ST}), we start with the generating functional of field correlators
in the Standard frame yet switch to the Gravic frame in the intermediate steps, wherein we compute the effective action ${\tilde{S}}_{eff}\left[\phi, \tilde{\psi}, \upsilon, \tilde{g}; \Lambda_{\texttt{EW}} \right]$. This effective field theory, however, is not the correct one for gravity to attract in the way it ought to, for $SM$ fields to interact in the way they ought to, and for Higgs field to be a true real scalar field as it ought to. Therefore, we map this effective action back to the Standard frame of long-wavelength fields via the conformal transformation
\begin{eqnarray}
\label{conf-trans-back}
\tilde{g}_{\mu \nu} = \left(\frac{\phi}{M_c}\right)^{2} \hat{g}_{\mu\nu}\,,\; \tilde{\psi} =  \left(\frac{\phi}{M_c}\right)^{-3/2} \hat{\psi}
\end{eqnarray}
which are, in structure, similar to inverse of the conformal transformations in (\ref{conf-trans}) except that the conformal factor here involves the long-wavelength Higgs $\phi$ not the complete one $\Phi$. The effective action thus obtained,  ${\hat{S}}_{eff}\left[\phi, \hat{\psi}, \upsilon, \hat{g}\right]$, will exhibit strikingly different properties compared to ${{S}}_{eff}\left[\phi, {\psi}, \upsilon, {g}; \Lambda_{\texttt{EW}}\right]$ in (\ref{part-func-ST}) in regard to the UV sensitivities of the model parameters. All these steps are summarized by the reduction process \cite{eff-action,eff-action2}
\begin{eqnarray}
\label{part-func-GR}
{Z} &=& \Fint \left[{\mathfrak{D}} \Phi\,  {\mathfrak{D}} \overline{{\Psi}}\, {\mathfrak{D}} {{\Psi}}\, {\mathfrak{D}} V_{\mu}\right]_{{g}} e^{{i} {{S}}\left[\Phi, {\Psi}, V, {g}\right]}\nonumber\\
&\xrightarrow{\text{Eq.}(\ref{conf-trans})}&\Fint \left[{\mathfrak{D}} \Phi\,  {\mathfrak{D}} \overline{\tilde{\Psi}}\, {\mathfrak{D}} {\tilde{\Psi}}\, {\mathfrak{D}} V_{\mu}\right]_{\tilde{g}} e^{{i} \left\{ {\tilde{S}}\left[\Phi, \tilde{\Psi}, V, \tilde{g}\right] + A_{\text{Eq.}(\ref{conf-trans})}\left[\Phi,\tilde{g}; \Lambda_{\texttt{UV}}\right]\right\} }\nonumber\\
&\xequal{{\text{Eq.}(\ref{fluctuations})}}&  \Fint \left[{\mathfrak{D}} \phi\,  {\mathfrak{D}} \overline{\tilde{\psi}}\, {\mathfrak{D}} {\tilde{\psi}} {\mathfrak{D}} \upsilon_{\mu}\right]_{\tilde{g}}\nonumber\\ &\times&
\left[{\mathfrak{D}}\left(\delta \phi\right)  {\mathfrak{D}}(\overline{\delta \tilde{\psi}}) {\mathfrak{D}}({\delta \tilde{\psi}}) {\mathfrak{D}}\left(\delta \upsilon_{\mu}\right)\right]_{\tilde{g}} e^{{i} \left\{ {\tilde{S}}\left[\Phi, \tilde{\Psi}, V, \tilde{g}\right] + A_{\text{Eq.}(\ref{conf-trans})}\left[\Phi,\tilde{g}; \Lambda_{\texttt{UV}}\right]\right\} }\nonumber\\
&=& \Fint \left[{\mathfrak{D}} \phi\,  {\mathfrak{D}} \overline{\tilde{\psi}}\, {\mathfrak{D}} {\tilde{\psi}}\, {\mathfrak{D}} \upsilon_{\mu}\right]_{\tilde{g}} e^{{i}\left\{ {\tilde{S}}_{eff}\left[\phi, \tilde{\psi}, \upsilon, \tilde{g}\right] + A_{\text{Eq.}(\ref{conf-trans})}\left[\phi,\tilde{g}; \Lambda_{\texttt{UV}}\right]\right\}}\nonumber\\
&\xrightarrow{\text{Eq.}(\ref{conf-trans-back})}& \Fint \left[{\mathfrak{D}} \phi\,  {\mathfrak{D}} \overline{\hat{\psi}}\, {\mathfrak{D}} {\hat{\psi}}\, {\mathfrak{D}} \upsilon_{\mu}\right]_{\hat{g}} e^{{i} \left\{ {\hat{S}}_{eff}\left[\phi, \hat{\psi}, \upsilon, \hat{g}\right] + A_{\text{Eq.}(\ref{conf-trans})}\left[\phi,\hat{g}; \Lambda_{\texttt{UV}}\right] + A_{\text{Eq.}(\ref{conf-trans-back})}\left[\phi,\hat{g}; \Lambda_{\texttt{EW}}\right]\right\}}
\end{eqnarray}
wherein $A_{\text{Eq.}(\ref{conf-trans})}\left[\Phi,\tilde{g}; \Lambda_{\texttt{UV}}\right]$ and $A_{\text{Eq.}(\ref{conf-trans-back})}\left[\phi,\hat{g}; \Lambda_{\texttt{EW}}\right]$ are the action functionals
induced by the anomalies resulting from the indicated conformal transformations. The anomalies originate from the changes in the
functional integration measures due to the conformal transformations (\ref{conf-trans}) and (\ref{conf-trans-back}).
As explicitly indicated in (\ref{part-func-ST}) and (\ref{part-func-GR}), functional integration measures depend on the metric,
and hence, they all change (on top of the changes in the fermion fields due to (\ref{conf-trans})) under the  conformal transformations (\ref{conf-trans}) and (\ref{conf-trans-back}). For instance, ${\mathfrak{D}} \phi$ in (\ref{part-func-ST}) is actually a tensor density because it involves $\left(- {g}\right)^{1/4}$ times $\phi$ \cite{fujikawa}. Direct calculation gives \cite{geometrical3}
\begin{eqnarray}
\label{anomal-action}
A_{\text{Eq.}(\ref{conf-trans})}\left[\phi,\hat{g}; \Lambda_{\texttt{UV}}\right] + A_{\text{Eq.}(\ref{conf-trans-back})}\left[\phi,\hat{g}; \Lambda_{\texttt{EW}}\right] &=&  \int d^{4}x \sqrt{-\hat{g}}\,  \frac{1}{(4\pi)^2} \left( 1+ 3 n_1 + 2 n_1^0 + 2 n_{1/2}\right) \nonumber\\ &\times& \left[ \Lambda_{\texttt{EW}}^4 - \frac{\Lambda_{\texttt{UV}}^4}{M_c^4} \phi^4  \right] \log\frac{\phi}{M_c}
\end{eqnarray}
where we have neglected subleading curvature effects, and employed the momentum cutoffs $\Lambda_{\texttt{UV}}$ for
$A_{\text{Eq.}(\ref{conf-trans})}\left[\Phi,\tilde{g}; \Lambda_{\texttt{UV}}\right]$, and $\Lambda_{\texttt{EW}}$ for $A_{\text{Eq.}(\ref{conf-trans-back})}\left[\phi,\hat{g}; \Lambda_{\texttt{EW}}\right]$.

The next step is the construction of the effective action ${\hat{S}}_{eff}\left[\phi, \hat{\psi}, \upsilon, \hat{g}; \Lambda_{\texttt{EW}} \right]$ in (\ref{part-func-GR}). For this purpose, it is necessary to first compute the Gravic frame's effective action ${\tilde{S}}_{eff}\left[\phi, \tilde{\psi}, \upsilon, \tilde{g}; \Lambda_{\texttt{EW}}\right]$. It is convenient to work with the canonical scalar
\begin{eqnarray}
d\aleph = \sqrt{K(\Phi)} d\Phi
\end{eqnarray}
defined through (\ref{changes2}). Using $\aleph = \chi + \delta \chi$ in (\ref{changes2}), and expanding it up to the quadratic order in the quantum fluctuation $\delta \chi$, we get the quadratic action
\begin{eqnarray}
\label{quad_F}
\Delta {\tilde{S}}_{\phi} &=& \frac{1}{2} \int d^{4} x\, \sqrt{-\tilde{g}}\, {\delta \chi} \left( - \tilde{\Box} + {\mathcal{M}}_{\phi}^2\right) {\delta \chi}
\end{eqnarray}
where
\begin{eqnarray}
\label{mass2-of-f}
{\mathcal{M}}^2_{\phi}= A_2(\phi)  M_{Pl}^2 \phi^{-2} {\tilde{{R}}}({\tilde{g}}) - A_2(\phi) m_{H}^2 M_{c}^2 \phi^{-2} - A_4 (\phi) V_0 M_{c}^2 \phi^{-4}
\end{eqnarray}
is the mass-squared of $\delta \chi$. The two new functions in it
\begin{eqnarray}
\label{quad-quart-func}
A_2(\phi) &=& \frac{1}{\left(6 M_{Pl}^2 \phi^{-2} -1\right)} - \frac{1}{\left(6 M_{Pl}^2 \phi^{-2} -1\right)^2}\,,\nonumber\\
A_4(\phi) &=& \frac{12}{\left(6 M_{Pl}^2 \phi^{-2} -1\right)} - \frac{4}{\left(6 M_{Pl}^2 \phi^{-2} -1\right)^2}\,,
\end{eqnarray}
necessarily involve the kinetic function $K(\phi)$. It is understood that $\phi \equiv \phi(\chi)$ everywhere in (\ref{mass2-of-f}) and (\ref{quad-quart-func}). The scale of ${\mathcal{M}}^2_{\phi}$ is determined by $m_H^2$ and $V_0$. From the beginning, as the defining scale,
we take $m_H^2$ to lie at the electroweak scale. On the other hand, the vacuum energy $V_0$ is unknown; it is not calculable and
can take any value up to $M_{Pl}^4$. Therefore, there are essentially two distinct regimes for the squared-mass of $\delta \chi$:
\begin{itemize}
\item Either ${\mathcal{M}}^2_{\phi} \sim \Lambda_{\texttt{EW}}$ for which $|V_0|$ can be at most $\Lambda_{\texttt{EW}}^4$,

\item Or ${\mathcal{M}}^2_{\phi} \gg \Lambda_{\texttt{EW}}$ in which case $|V_0|$ can be as large as $\Lambda_{\texttt{UV}}^4\sim M_{Pl}^4$.
\end{itemize}
Though the first one seems more natural for electroweak theory, both regimes are acceptable and gives rise to no technical or
conceptual problems in the analysis. After going back to the Standard frame, $V_0$ will again be the tree-level  vacuum energy
density.

Coming to vector and spinor fluctuations, they are found to possess the quadratic action
\begin{eqnarray}
\label{quad_Psi}
\Delta {\tilde{S}}_{\psi-\upsilon} &=& \int d^{4} x\, \sqrt{-\tilde{g}}\, \Bigg\{ \overline{\delta{\tilde \psi}}\left( - \slashed{\nabla} - \frac{1}{\sqrt{2}} \lambda_{\Psi} M_c \right) \delta\tilde{\psi} - i g \left(\overline{\delta{\tilde \psi}} {\delta{\slashed{\upsilon}}} \tilde{\psi} + \overline{{\tilde \psi}} \delta{\slashed{\upsilon}} \delta\tilde{\psi}\right) \nonumber\\
&+& \delta\upsilon_{\mu} \left( \Box \tilde{g}^{\mu \nu} - \tilde{R}^{\mu \nu} - \lambda_V M_c^2 \tilde{g}^{\mu\nu} \right) \delta\upsilon_{\nu}\Bigg\}
\end{eqnarray}
where flavor and gauge indices are suppressed \cite{tye,coleman-weinberg}. The $\delta{\tilde{\psi}}-\delta\upsilon_{\mu}$ mixing, the second term in the action, gives no significant contribution because of the $1/M_c$ suppression it suffers compared to the vector and fermion contributions.

For constructing the effective action ${\tilde{S}}_{eff}\left[\phi, \tilde{\psi}, \upsilon, \tilde{g}; \Lambda_{\texttt{EW}}\right]$, we integrate over $\delta\chi$, $\delta\upsilon_{\mu}$, $\delta\tilde{\psi}$ in the quadratic actions (\ref{quad_F}), (\ref{quad_Psi}),
in the vertices (\ref{generic-couplings}), and in the kinetic terms of the fields. All these corrections constitute the $\widetilde{SM}$ effective field theory at the electroweak scale.

\subsubsection{Effective Potential}
\label{sec321}
Among all corrections, the most crucial one is the effective potential which, upon transforming back to  the Standard frame via (\ref{conf-trans-back}), assumes the form
\begin{eqnarray}
\label{1loop_pot-ST-back}
\widehat{V}_{eff}\left(\phi, \hat{g}; \Lambda_{\texttt{EW}}\right) = \overline{V_0}  + \frac{1}{2} \overline{m_{H}^2}
\phi^{2} + \frac{1}{4} \overline{\lambda_{H}}\phi^4 - \frac{1}{2} \left(\overline{M_{Pl}^2} -
\overline{\lambda_R} \phi^2 \right) {\hat{R}}({\hat{g}})
\end{eqnarray}
which is to replace the effective potential $V_{eff}(\phi,g; \Lambda_{\texttt{EW}})$ in (\ref{1loop_pot-ST-back}) computed directly in the Standard frame. Here, the renormalized dimensionful couplings are given by
\begin{eqnarray}
\label{renormed_Mpl-GR}
{{\overline{M_{Pl}^2}}} &=& M_{Pl}^2 \left( 1 - \frac{A_2(\phi)}{(4\pi)^2} \left[ \frac{{\Lambda_{\texttt{UV}}^2}}{M_{c}^2} - A_{24}(\phi) \log \frac{\Lambda_{\texttt{EW}}^2}{\Lambda_{\texttt{UV}}^2} \right]\right),\\
\label{renormed_Mphi-GR}
{{\overline{m_{H}^2}}} &=& m_{H}^2 \left( 1 - \frac{A_2(\phi)}{(4\pi)^2} \left[ \frac{{\Lambda_{\texttt{UV}}^2}}{M_{c}^2} - \frac{1}{2} A_{24}(\phi) \log \frac{\Lambda_{\texttt{EW}}^2}{\Lambda_{\texttt{UV}}^2} \right]\right),\\
\label{renormed_V0-GR}
{{\overline{V_0}}} &=& V_0 \left( 1 - \frac{A_4(\phi)}{(4\pi)^2} \left[ \frac{{\Lambda_{\texttt{UV}}^2}}{M_{c}^2} - \frac{1}{2} A_{24}(\phi) \log \frac{\Lambda_{\texttt{EW}}^2}{\Lambda_{\texttt{UV}}^2} \right]\right),
\end{eqnarray}
which explicitly involve the long-wavelength Higgs field $\phi$ via $A_{2}(\phi)$, $A_4(\phi)$ and their combination
\begin{eqnarray}
A_{24}(\phi) = A_2(\phi) m_{H}^2 \phi^{-2} + A_4(\phi) V_0 \phi^{-4}\,.
\end{eqnarray}

The dimensionless couplings in (\ref{1loop_pot-ST-back}) renormalize as
\begin{eqnarray}
\label{renormed_zeta-GR}
{{\overline{\lambda_R}}} &=& - \frac{1}{(4\pi)^2} \left[ \left(\frac{1}{6} - \frac{n_1^0}{6} - \frac{n_1}{4} + \frac{n_{1/2}}{3}\right) \frac{{\Lambda_{\texttt{UV}}^2}}{M_{c}^2} - \frac{1}{6} \left( \frac{3}{2} \lambda_V - \lambda_{\psi}^2 + A_{24}(\phi)\right)  \log \frac{\Lambda_{\texttt{EW}}^2}{\Lambda_{\texttt{UV}}^2}\right],\\
\label{renormed-lamphi-GR}
\overline{\lambda_{H}} &=& \lambda_{H} + \frac{1}{(4 \pi)^2}\left[ (n_F-n_B) \frac{\Lambda_{\texttt{UV}}^4}{M_{c}^4} + \left( 6 \lambda_V - 4 \lambda_{\psi}^2 \right) \frac{{\Lambda_{\texttt{UV}}^2}}{M_{c}^2}  + \left(3 \lambda_V^2 - \lambda_{\psi}^4\right)  \log \frac{\Lambda_{\texttt{EW}}^2}{\Lambda_{\texttt{UV}}^2}\right],
\end{eqnarray}
where $\overline{\lambda_{H}}$ is seen to receive all sort of corrections: quartic, quadratic and logarithmic.

A short glance at the renormalization properties of parameters reveals striking differences between the effective
potentials (\ref{1loop-pot-ST}) and (\ref{1loop_pot-ST-back}). The reasons for and results of the differences
lie at the heart of the mechanism being proposed. It can thus prove useful to dwell on certain salient features:
\begin{itemize}
\item {\bf Dimensionful parameters renormalize multiplicatively.} The dimensionful parameters are renormalized multiplicatively in manifest contrast to their additive renormalizations in the Standard frame (compare (\ref{renormed_Mpl-GR}), (\ref{renormed_Mphi-GR}) and (\ref{renormed_V0-GR}) with the equations (\ref{renormed_Mpl-ST}), (\ref{renormed_Mphi-ST}) and (\ref{renormed_V0-ST})). Multiplicative renormalization guarantees that, any of these mass parameters, if vanishes at the tree level, stays  vanishing at all orders. This is similar to the renormalization of fermion masses. In our case, it is scale invariance, not chiral invariance, which protects the dimensionful parameters. This is the case because, not the action in the Standard frame, but the action in the Gravic frame exhibits a scaling symmetry ($k$ is a parameter)
    \begin{eqnarray}
    \label{simetri1}
    {\tilde{S}}\left[k\phi,\tilde{g}; k^2 {M_{Pl}^2}, k^2 {m_{H}^2}, k^4 {V_0}\right] =  {\tilde{S}}\left[\phi,\tilde{g}; {M_{Pl}^2}, {m_{H}^2}, {V_0} \right]
    \end{eqnarray}
    as is seen from (\ref{changes}) and (\ref{changes2}). As a consequence of this, ${\mathcal{M}}^2_\phi$ in the quadratic action (\ref{quad_F}) obeys the same scaling relation
    \begin{eqnarray}
    \label{simetri2}
    {\mathcal{M}}^2_{\phi}\left(k \phi; k^2 {M_{Pl}^2}, k^2 {m_{H}^2}, k^2 {V_0}\right) = {\mathcal{M}}^2_{\phi}\left(\phi; {M_{Pl}^2}, {m_{H}^2}, {V_0}\right)
    \end{eqnarray}
     which can be read off from (\ref{mass2-of-f}). This scaling symmetry is a by-product of the conformal transformations (\ref{conf-trans}), and  arises because $\phi$ dresses the fundamental parameters $M_{Pl}$, $m_{H}^{2}$ and $V_0$. Being singlets under this scaling transformation, $M_{c}$ and ${\tilde{R}}(\tilde{g})$ can contribute to the renormalization process only by themselves (like $\lambda_{\phi} M_{c}^4$ term). Therefore, ${\mathcal{M}}^2_{\phi}$ cannot contain terms like $const\times M_{c}^2$ or $const\times M_{c}^4 \phi^{-2}$ or $const\times \phi^2$ because of the scaling symmetry. This symmetry  is actually what forbids the fundamental parameters to
     receive additive quantum corrections. Indeed, for instance, a term like $const\times M_{c}^4 \phi^{-2}$ would generate a quadratically divergent additive term as in the Standard frame. Consequently, dimensionful parameters ought to renormalize multiplicatively.

\item {\bf Dimensionful parameters receive tiny quantum corrections.} In complete contrast to the Standard frame results in (\ref{renormed_Mpl-ST}), (\ref{renormed_Mphi-ST}) and (\ref{renormed_V0-ST}), here quantum corrections to dimensionful parameters turn out to be their diminutive fractions. This is seen from the leading parts of (\ref{renormed_Mpl-GR}), (\ref{renormed_Mphi-GR}) and (\ref{renormed_V0-GR})
     \begin{eqnarray}
\label{renormed_Mpl-GR-approx}
{{\overline{M_{Pl}^2}}} &\approx& M_{Pl}^2 \left( 1 - \frac{1}{(4\pi)^2} \left[ \frac{\phi^2}{6 M_{Pl}^2} \frac{{\Lambda_{\texttt{UV}}^2}}{M_{c}^2} - \left( \frac{V_0}{3 M_{Pl}^4}+ \frac{ m_{H}^2 \phi^2} {36 M_{Pl}^4} \right) \log \frac{\Lambda_{\texttt{EW}}^2}{\Lambda_{\texttt{UV}}^2}\right]\right),\\
\label{renormed_Mphi-GR-approx}
{{\overline{m_{H}^2}}} &\approx& m_{H}^2 \left( 1 - \frac{1}{(4\pi)^2} \left[ \frac{\phi^2}{6 M_{Pl}^2} \frac{{\Lambda_{\texttt{UV}}^2}}{M_{c}^2} + \left( \frac{V_0}{6 M_{Pl}^4} - \frac{ m_{H}^2 \phi^2} {72 M_{Pl}^4} \right) \log \frac{\Lambda_{\texttt{EW}}^2}{\Lambda_{\texttt{UV}}^2}\right]\right),\\
\label{renormed_V0-GR-approx}
{{\overline{V_0}}} &\approx& V_0 \left( 1 - \frac{1}{(4\pi)^2} \left[ \frac{2 \phi^2}{M_{Pl}^2} \frac{{\Lambda_{\texttt{UV}}^2}}{M_{c}^2} - \left( \frac{2 V_0}{M_{Pl}^4}+ \frac{ m_{H}^2 \phi^2} {6 M_{Pl}^4} \right) \log \frac{\Lambda_{\texttt{EW}}^2}{\Lambda_{\texttt{UV}}^2}\right]\right),
\end{eqnarray}
which are obtained by an expansion in powers of $\phi/M_{Pl}$ up to ${\cal{O}}\left(1/M_{Pl}^{4}\right)$. These corrections become tinier and tinier as $M_{c}$ takes larger and larger values. In fact, to an excellent approximation, one can take all three mass scales, ${{\overline{M_{Pl}^2}}}$, ${{\overline{m_{H}^2}}}$, ${{\overline{V_0}}}$, as staying put at their tree-level values.

\item {\bf Dimensionless parameters renormalize additively.} The dimensionless couplings renormalize additively by receiving not only logarithmic corrections as in the Standard frame (see the equations  (\ref{renormed_zeta-ST}) and  (\ref{renormed-lamphi-ST})) but  also quartic and quadratic corrections. This property stems from modifications in their roles in switching to the Gravic frame, and hence,  quantum corrections to dimensionless couplings can be more sizeable than those to dimensionful ones. For instance,  the quartic coupling $\overline{\lambda_{H}}$, as given in (\ref{renormed-lamphi-GR}), can exceed unity or become negative if corrections are sizeable. This, however, is not the case. The reason is that $\Lambda_{\texttt{UV}}$ is at most $M_{Pl}$ and $M_c\gtrsim M_{Pl}$ according to the unitarity bound (\ref{McMpl}). Thus $\Lambda_{\texttt{UV}}$ lies well below $M_c$ to diminish the quantum corrections to $\lambda_H$ so that Higgs potential stays away from vacuum unstability and non-perturbativity regimes.

\item {\bf Anomalies cause disruptions.} The effective potential (\ref{1loop_pot-ST-back}) does not include the contributions of the anomaly-induced action (\ref{anomal-action}). Their inclusion shifts effective potential by
    \begin{eqnarray}
\label{1loop_pot-ST-shift}
\delta \widehat{V}_{eff}\left(\phi, \hat{g}; \Lambda_{\texttt{EW}}\right) = \delta \overline{V_0} + \frac{1}{4} \delta\overline{\lambda_{H}}\, \phi^4
\end{eqnarray}
with
\begin{eqnarray}
\delta \overline{V_0}  &=& - \frac{1}{(4\pi)^2} \left( 1 + {3} n_1 + 2 n_1^0 + 2 n_{1/2}\right)  \Lambda_{\texttt{EW}}^4 \log \frac{\phi}{M_c}\\
\delta\overline{\lambda_{H}} &=& \frac{1}{(4\pi)^2} \left( 4 + 12 n_1 + 8 n_1^0 + 8 n_{1/2}\right) \frac{\Lambda_{\texttt{UV}}^4}{M_{c}^4} \log \frac{\phi}{M_c}\,.
\end{eqnarray}
Obviously, $\delta\overline{\lambda_{H}}$ shifts the Higgs quartic coupling with a similar size as the other contributions in (\ref{renormed-lamphi-GR}). Also, $\overline{\lambda_{H}}$ develops a $\phi$ dependence which is rather mild since $\log \frac{\phi}{M_c}\simeq \log\frac{\Lambda_{\texttt{EW}}}{\Lambda_{\texttt{UV}}}$.

The main effect of $\delta \overline{V_0}$ is to break the scaling law in (\ref{simetri2}). This is expected since quantum effects break scale invariance. In consequence, conformal protection on $\overline{V_0}$ is lifted. Nevertheless, this breaking is soft in that the added correction
is ${\mathcal{O}}\left( \Lambda_{\texttt{EW}}^4 \right)$ not ${\mathcal{O}}\left( \Lambda_{\texttt{UV}}^4 \right)$ or some intermediate-scale energy density. In spite of this, having $\overline{V_0}\sim {\mathcal{O}}\left( \Lambda_{\texttt{EW}}^4 \right)$ is in any case a disaster since, observationally, $\overline{V_0}$ must read $m_{\nu}^{4}$ \cite{ccp-exp}.

\end{itemize}

\subsubsection{Effective Vertices}
\label{sec322}
Effective potential is only part of the story. The effective field theory described by ${\hat{S}}_{eff}\left[\phi, \hat{\psi}, \upsilon, \hat{g}; \Lambda_{\texttt{EW}} \right]$ has all its couplings and fields renormalized at the electroweak scale. In a full renormalization programme all these corrections are to be computed to form the $\widehat{SM}$ quantum field theory below the electroweak scale $\Lambda_{\texttt{EW}}$. Nevertheless, for revealing salient features of the Gravic frame (especially the matter-Higgs decoupling), it should suffice to analyze certain representative observables. Gauge invariance and finite UV cutoff control the perturbation series. Here we focus on  the Higgs interactions in (\ref{generic-couplings}). As required by (\ref{part-func-GR}), it is necessary to first compute the quantum corrections in the Gravic frame by considering the interactions in (\ref{changes}). Then, by applying the conformal transformations (\ref{conf-trans-back}), we map the vertices back to the Standard frame, to find
\begin{eqnarray}
\label{generic-couplings-renormed}
- \frac{1}{2} \overline{\lambda_V} \phi^2 \hat{g}^{\mu \nu} \upsilon_{\mu} \upsilon_{\nu} - \frac{1}{\sqrt{2}} \overline{\lambda_{\Psi}} \phi \overline{\hat{\psi}} \hat{\psi}\,.
\end{eqnarray}
In view of the transformations (\ref{changes}) and (\ref{trans-expect}), the tree-level vertices in (\ref{generic-couplings}) become mass terms. Therefore, the renormalized couplings $\overline{\lambda_V}$ and $\overline{\lambda_{\Psi}}$ originate from the vector boson and fermion self-energy diagrams in the Gravic frame. For instance, $\frac{1}{2} \lambda_V \Phi^2 g^{\mu \nu} V_{\mu} V_{\nu}$ in (\ref{generic-couplings}) become $\frac{1}{2} \lambda_V M_c^2 \tilde{g}^{\mu \nu} V_{\mu} V_{\nu}$ after the conformal transformation (\ref{conf-trans}). In this frame, loops of $\delta\psi$ (and also the loops of $\delta\upsilon_{\mu}$ for non-Abelian gauge fields as discussed in Sec. \ref{sec313}) induce the self-energy of $\upsilon_{\mu}$, and this self-energy term generates $\overline{\lambda_V}$ upon the transformation (\ref{conf-trans-back}):
\begin{eqnarray}
\label{eff-lam-V}
 \overline{\lambda_V} =  {\lambda_V} - \frac{g^2}{8\pi^2}\frac{\frac{{\Lambda_{\texttt{UV}}^4}}{M_{c}^4}}{\frac{\lambda_{\Psi}^2}{2} +  \frac{{\Lambda_{\texttt{UV}}^2}}{M_{c}^2}}
\end{eqnarray}
which renormalizes the gauge coupling with a strength depending on how small ${\Lambda_{\texttt{UV}}}$ with respect to $M_c$. The renormalization above is due to the fermion loops, only.  In a similar fashion, the fermion self-energy in the Gravic frame induces $\overline{\lambda_{\Psi}}$ in Standard frame
\begin{eqnarray}
\label{eff-lam-Psi}
 \overline{\lambda_{\Psi}} =  {\lambda_{\Psi}}\left( 1 - \frac{g^2}{4\pi^2}\left[\frac{\lambda_V}{\lambda_V - \frac{\lambda_{\Psi}^2}{2}} \log \left( 1- \frac{1}{\lambda_V} \frac{{\Lambda_{\texttt{UV}}^2}}{M_{c}^2} \right) + \left( \lambda_V \leftrightarrow \frac{\lambda_{\Psi}^2}{2}\right)\right]\right)
\end{eqnarray}
which gives a logarithmic multiplicative renormalization of the Yukawa coupling $\lambda_{\Psi}$. There are other kinds of vertices as well. For instance, vector boson-fermion-fermion interaction exists in both Gravic and Standard frames, and receives a vertex-type correction in either.  Among all, the most anomalous ones are the Higgs-matter couplings. They totally vanish in the Gravic frame, and cannot be induced at any order of perturbation theory. They are generated when one switches to Standard frame via (\ref{conf-trans-back}), and quantum corrections to such couplings
are obtained within the electroweak theory with the UV scale $\Lambda_{\texttt{EW}}$.

The meaning of a Higgs interaction term in the Standard frame and in the Gravic frame can be strikingly different, as clearly shown by the ways $\overline{\lambda_V}$ and $\overline{\lambda_{\Psi}}$ are generated. For instance, $\overline{\lambda_V}$ in (\ref{eff-lam-V}) results from self-energy of a massive vector boson. This dictates how this renormalized coupling should depend on tree-level parameters. Nevertheless, as is clearly seen from (\ref{generic-couplings-renormed}), after returning to $\widehat{SM}$ Standard frame, the same coupling becomes  Higgs-Higgs-Vector-Vector quartic coupling. It then acts as `tree-level coupling' (for $\upsilon_{\mu}$ self-energy,  $h\upsilon\upsilon$ coupling as well as $hh\upsilon\upsilon$ coupling after electroweak breaking) for renormalization comprising quantum fluctuations with frequencies up to ${\Lambda_{\texttt{EW}}}$. The situation would be different if renormalization were carried out entirely in the Standard frame ( in which case Higgs sector suffers from the gauge hierarchy problem). In that case, one starts with $\upsilon_{\mu}$ mass term as well as $hVV$ and $hhVV$ couplings, all being fixed by the tree-level coupling $\lambda_V$, and renormalize each coupling by taking into account quantum fluctuations (including those of the Higgs field) for frequencies up to ${\Lambda_{\texttt{UV}}}$. Therefore, the two approaches differ from each other for frequencies ranging from ${\Lambda_{\texttt{EW}}}$ up to ${\Lambda_{\texttt{UV}}}$. In fact, in this energy range differences between the two models
are similar to those found in Higgsless models \cite{renorm1}. Nevertheless, after constructing the effective field theory at the electroweak scale, the two approaches receive identical loop contributions (for frequencies up to  ${\Lambda_{\texttt{EW}}}$). Either renormalization procedure can be implemented into a global precision analysis of the electroweak observables \cite{ew-precision}.

\subsubsection{An Application: $h \rightarrow \gamma \gamma$ Decay}
\label{sec323}
The Higgs field, having no electric and color charges, do not couple to
(strictly massless) photon and gluon at the tree level. This holds both in
Gravic and Standard frames. In the Gravic frame, where Higgs does not
couple to any single matter field, Higgs is prohibited to
couple to photon and gluon even at the quantum level. In Standard
frame, however, Higgs   develops couplings to photon and
gluon at the loop level. These quantum-induced interactions
embody various interesting features of the renormalization
procedure we are pursuing. Therefore, we prefer to illustrate workings of the
formalism by outlining the $h \rightarrow \gamma \gamma$ decay.

\begin{enumerate}
\item Our framework is set by the $\widehat{SM}$ quantum field theory described by the effective action
${\hat{S}}_{eff}\left[\phi, \hat{\psi}, \upsilon, \hat{g}; \Lambda_{\texttt{EW}}\right]$. It is
the effective action in the last line of (\ref{part-func-GR}), and must be interpreted together
with the indicated anomaly contributions.

\item By minimizing the effective potential in (\ref{1loop_pot-ST-back}) plus the anomaly contribution in (\ref{1loop_pot-ST-shift}),
one determines the Higgs VEV:
\begin{eqnarray}
\label{vev-higgs}
\langle \phi^2 \rangle = - \frac{ \overline{m_{H}^2}}{\left(\overline{\lambda_{H}} + \delta\overline{\lambda_{H}}\right)}
\end{eqnarray}
where tiny multiplicative renormalization of $m_H^2$ given in (\ref{renormed_Mphi-GR-approx}) guarantees that $\overline{m_{H}^2} < 0$ if ${m_{H}^2} < 0$. The corrections to quartic coupling is well under control because $\Lambda_{\texttt{UV}}$ is well below $M_c$. This quantum stability of $\langle \phi^2 \rangle$, which was no way reachable with (\ref{renormed_Mphi-ST}) and (\ref{renormed-lamphi-ST}), is the main point of the renormalization programme we are pursuing.

\item The Higgs VEV (\ref{vev-higgs}) changes the net vacuum energy. Indeed, with non-vanishing $\langle \phi \rangle$, the total vacuum energy becomes
\begin{eqnarray}
\overline{V}_{tot}\left(\Lambda_{\texttt{EW}}\right) &=& \overline{V_0} + \delta \overline{V_0} - \frac{1}{4} {\left(\overline{\lambda_{H}} + \delta\overline{\lambda_{H}}\right)} \langle \phi^2 \rangle^2\,.
\end{eqnarray}
This energy density, unlike its Standard frame counterpart (\ref{1loop-pot-ST}), takes a value right at the electroweak scale. For energies below the neutrino mass scale, the effective potential $\overline{V}_{tot}\left(m_{\nu}\right)$ must observationally have a value close to $m_{\nu}^{4}$. Satisfying this requires a fine cancellation between $V_0$ and the other pieces up to some sixty digits \cite{ccp-weinberg,nima}. Therefore, despite the UV insensitivity brought by the Gravic frame, the cosmological constant problem stands still with its entire perplexity. Nevertheless,  the problem metamorphoses to become a problem of the electroweak scale not that of the higher energy scales though the $SM$ is taken to be valid up to the Planck scale.

\item Having $\langle \phi^2 \rangle$ fixed to the electroweak scale without any particular fine-tuning guarantees that the Fermi scale
and particle masses generated all read at the electroweak scale \cite{peskin}. The Higgs boson
    \begin{eqnarray}
    h = \phi - \langle \phi \rangle
    \end{eqnarray}
is in the particle spectrum with its mass-squared
\begin{eqnarray}
\label{m-higgs}
m_h^2 = - 2 \overline{m_{H}^2}
\end{eqnarray}
fixed to the electroweak scale. Nevertheless, this mass must further be corrected since it does not encapsulate
yet quantum fluctuations with frequencies below ${\Lambda_{\texttt{EW}}}$. In other words, within the $\widehat{SM}$ quantum field theory described by ${\hat{S}}_{eff}\left[\phi, \hat{\psi}, \upsilon, \hat{g}; \Lambda_{\texttt{EW}}\right]$, quantum
corrections to scattering processes and model parameters are yet to be included. In other words, interactions in the $\widehat{SM}$ are to be further renormalized with the UV scale $\Lambda_{\texttt{EW}}$. The UV scale is now the electroweak scale itself. Therefore, the Higgs boson mass in (\ref{m-higgs}), for instance, is renormalized as \cite{bardin-passarino}
\begin{eqnarray}
\label{m-higgs-ren}
\left({m_h^2}\right)_{ren} &=& m_h^2 + \frac{1}{(4\pi)^2} \Bigg[3 m_W^2 + \frac{m_Z^2}{2 c_W^2}
 + \frac{9 m_h^4}{8 m_W^2} - \frac{6 m_t^4}{m_W^2}
+ 6 m_W^2 \log \frac{m_W^2}{\Lambda_{\texttt{EW}}^2}\nonumber\\ &+& \left( \frac{3}{2} + \frac{1}{2 c_W^2}\right) m_Z^2 \log\frac{m_Z^2}{\Lambda_{\texttt{EW}}^2} + \left( \frac{3}{4} + \frac{9 m_h^2}{8 m_W^2}\right) m_h^2 \log \frac{m_h^2}{\Lambda_{\texttt{EW}}^2}
\nonumber\\&-& \frac{9 m_t^2}{m_W^2} m_t^2 \log \frac{m_t^2}{\Lambda_{\texttt{EW}}^2}
+ 3 \left( 1 + \frac{1}{2 c_W^2} + \frac{m_h^2}{4 m_W^2} - \frac{m_t^2}{m_W^2} \right) {\Lambda_{\texttt{EW}}^2}\Bigg]
\end{eqnarray}
where $m_W^2= \frac{1}{4} \overline{\lambda_W} \langle \phi^2 \rangle$, $m_Z^2= \frac{1}{4} \overline{\lambda_Z} \langle \phi^2 \rangle$
and $m_t = \frac{1}{\sqrt{2}} \overline{\lambda_t} \langle \phi \rangle$ with $\overline{\lambda_W}$, $\overline{\lambda_Z}$, $\overline{\lambda_t}$ being the couplings renormalized in the Gravic frame as in (\ref{eff-lam-V}) and (\ref{eff-lam-Psi}). In this
expression the last term is particularly interesting. It is proportional to ${\Lambda_{\texttt{EW}}^2}$ and  heralds that the quadratic divergence is revived. This is not surprising because we are in the Standard frame of the conformal transformations (\ref{conf-trans-back}), and, as discussed in Sec. \ref{sec2},  quadratic divergences
naturally arise in this frame. This, however, does not spoil naturalness since the UV scale ${\Lambda_{\texttt{EW}}}$ is the electroweak scale itself. Nonetheless, ${\Lambda_{\texttt{EW}}}$ must lie sufficiently close to the Fermi scale as otherwise the little hierarchy problem  starts surfacing.

The Higgs boson couples to quarks, leptons and vector bosons via (\ref{generic-couplings-renormed}). The particle masses weigh right at the electroweak scale with gauge and Yukawa couplings given in (\ref{eff-lam-V}) and (\ref{eff-lam-Psi}). One, however, keeps in mind that these masses along with various couplings must still be corrected for quantum fluctuations of fields with frequencies below ${\Lambda_{\texttt{EW}}}$.

\item Higgs boson couples to two photons via $W$ boson and top quark loops. The diagrams, in the cutoff regularization method employed here, have recently been computed in \cite{cinli}. The amplitude is finite and independent of the UV scale ${\Lambda_{\texttt{EW}}}$. The decay rate agrees with known results. The main difference is that the masses and couplings of the fields correspond to those in (\ref{generic-couplings-renormed}) where quantum fluctuations with frequencies from ${\Lambda_{\texttt{EW}}}$ up to ${\Lambda_{\texttt{UV}}}$ are incorporated in the renormalized parameters (\ref{eff-lam-V}) and (\ref{eff-lam-Psi}). Needless to say, observation of $h \rightarrow \gamma \gamma$ at the LHC, alone, indicates that the $SM$ cannot be in the Gravic frame at the electroweak scale. On the other hand, possible
    non-observation of `new physics' would be an indication of the fact that the $SM$ can be in the Gravic frame beyond Fermi energies.

\end{enumerate}

\subsection{New Physics}
\label{sec33}
Throughout the text, emphasis has been put on the $SM$ fields. However, as already mentioned in Sec. \ref{sec311}, extensions of the $SM$ are well accommodated provided that the Higgs mass naturalness is maintained. To this end, possible extensions must form a CFT whose fields can acquire their masses and other dimensionful couplings, if they are to, from the Higgs mechanism through their couplings to the Higgs field. In this sense, the scale of new physics cannot be far from the electroweak scale. Their experimental verification, however, depends crucially on the strengths and structures of their couplings to the $SM$ fields.

\begin{itemize}
\item {\bf Sequential Models and Neutrinos.} There are different possibilities for `new physics'. One possibility pertains extra families of quarks and leptons. Another possibility concerns extra gauge groups (whose anomalies can be cancelled by introducing appropriate matter multiplets). These are the types of new physics models where extra matter directly interacts with the Higgs doublet. To this end, an interesting extension is provided by massive neutrinos \cite{neutrino}.  They are perfectly admissible if their masses are Dirac. If they are Majorana neutrinos, the ghosty Higgs of the Gravic frame decays into neutrinos and vacuum is destabilized with a rate depending on the right-handed neutrino mass scale. To the extent these decays are admissible the Majorana neutrinos might be accommodated \cite{demir,vilenkin}. In general, if new physics qualifies to be a CFT then, in the Gravic frame, the new fields, get decoupled from the Higgs, and hence, their radiative effects do not spoil the naturalness.

\item {\bf Extra Higgs Bosons and CP Violation.} As a more general extension of the $SM$, one can consider adding a second Higgs doublet (with the same hypercharge as $H$), say $H^{\prime}$, as in two-Higgs-doublet models \cite{2hdm}. In this case, the primary constraint is that $H^{\prime}$ must have no hard mass {\it i.e.} $m_{H^{\prime}} \equiv 0$. Its mass must originate solely from the $SM$ Higgs doublet via, for instance, the conformal coupling $\left(H^{\dagger}H\right) ({H^{\prime}}^{\dagger} H^{\prime})$. In Gravic frame, $\tilde{H^{\prime}}$ decouples from $H$, acquires a hard squared-mass proportional to $M_c^2$, and receives quantum corrections proportional to ${\Lambda_{\texttt{UV}}}^2$ to its squared-mass. In the Standard frame of (\ref{conf-trans-back}), as a long-wavelength field $\hat{H^{\prime}}$, it becomes massless again yet its various couplings get renormalized.

    The interaction potential of the two doublets
    \begin{eqnarray}
    && \frac{1}{2} m_H^2 H^{\dagger} H +  \frac{1}{4} \lambda_{H} \left(H^{\dagger}H\right)^2 + \frac{1}{4} \lambda_{H^{\prime}}  \left({H^{\prime}}^{\dagger}{H^{\prime}}\right)^2 + \frac{1}{2} \lambda_{HH^{\prime}} \left(H^{\dagger}H\right) ({H^{\prime}}^{\dagger} H^{\prime})\nonumber\\ &+& \frac{1}{2} {\lambda}_{1}
    \left(H^{\dagger}H^{\prime}\right) ({H^{\prime}}^{\dagger} H) + \frac{1}{2} \left[ \lambda_{2} \left(H^{\dagger}H^{\prime}\right)^2 + {\text{H.C.}}\right]\nonumber\\ &+& \frac{1}{2} \left[ {\lambda}_{3} \left(H^{\dagger}H\right) \left(H^{\dagger} H^{\prime}\right) + {\text{H.C.}}\right] + \frac{1}{2} \left[ {\lambda}_{4} \left({H^{\prime}}^{\dagger}H^{\prime}\right) \left(H^{\dagger} H^{\prime}\right) + {\text{H.C.}}\right]
    \end{eqnarray}
    possesses enough complex parameters ($\lambda_{2,3,4}$) to accommodate explicit CP violation. The spectrum then involves a charged Higgs boson, and three indefinite-CP neutral scalars which reduce to two CP-even and one CP-odd scalar in the CP-conserving limit. This extended Higggs sector, if any, can be discovered at the LHC experiments. The $H^{\prime}$  develops a mass after electroweak breaking in $\widehat{SM}$, and adds the aforementioned neutral and charged scalars to the spectrum \cite{2hdm}. This two-Higgs doublet model can realize electroweak baryogenesis depending on the nature of phase transition \cite{baryogen}.

\item {\bf The $SM$ Portals to Hidden Sector.}
It is highly conceivable that `new physics' can consist of fields which are singlets under the $SM$ gauge group. In other words, new physics can contain a singlet CFT sector which can, in general, consist of all possible fields and symmetries provided that there is no new mass scale involved. Actually, out of all these fields, only scalars and vectors directly communicate with the $SM$ fields. Therefore, as far as interactions with the $SM$ are concerned, the reactive part of the $SM$ singlet CFT sector reads as
\begin{eqnarray}
\label{lagran-singlet}
&&- \sqrt{-g}\Bigg[ \frac{a}{4}  g^{\alpha\mu} g^{\beta\nu} \Sigma^{-}_{\mu \nu} \Sigma^{-}_{\alpha \beta} + \frac{b}{4} g^{\alpha\mu} g^{\beta\nu} \Sigma^{+}_{\mu \nu} \Sigma^{+}_{\alpha \beta} + g^{\mu\nu} \left({\mathcal{D}}_{\mu} S\right)^{\dagger} \left({\mathcal{D}}_{\nu} S\right) + \lambda_S \left(S^{\dagger} S\right)^2\nonumber\\ && + \zeta_c  S^{\dagger} S R(g) + \lambda_{S1} S^{\dagger} S  g^{\mu\nu} \Sigma_{\mu}\Sigma_{\nu} \Bigg]
\end{eqnarray}
in the Standard frame. It is fully conformal-invariant for $\zeta_c = 1/6$ \cite{conf-trans}. Here, $S$ is a charged scalar (under some hidden gauge group), and $\Sigma_{\mu}$ is a non-gauge vector field with $\Sigma^{\pm}_{\mu\nu} = \nabla_{\mu}\Sigma_{\nu} \pm \nabla_{\nu} \Sigma_{\mu}$. This scalar-vector theory interacts with the $SM$ fields as
\begin{eqnarray}
\label{portals}
 - \sqrt{-g} \left[\lambda_{H0} \left(H^{\dagger} H\right) \left(S^{\dagger} S\right) +  \lambda_{H1} \left(H^{\dagger} H\right) g^{\mu \nu} \Sigma_{\mu} \Sigma_{\nu} + \lambda_{\Psi1} \overline{\Psi} \slashed{\Sigma} \Psi + \lambda_{B1} g^{\alpha\mu} g^{\beta\nu} B_{\alpha\beta} \Sigma^{-}_{\mu \nu}   \right]
\end{eqnarray}
wherein the $B_{\mu}$ is the hypercharge gauge field with the field strength tensor $B_{\mu \nu} = \partial_{\mu} B_{\nu} - \partial_{\nu} B_{\mu}$.
The spinor $\Psi$ counts gauge multiplets of all the $SM$ fermions ($SU(2)$ doublets of quarks and leptons as well as the right-handed leptons and quarks). Each term in (\ref{portals}) provides a specific channel for the singlet conformal new physics to interact with the $SM$. They each can be constrained by appropriate measurements. For instance, the fermion portal involves a $SM$-singlet non-gauge massless vector boson $\Sigma_{\mu}$, and LEP and Tevatron experiments must be rather sensitive to its contribution (that adds to the photon contribution).  Similar experimental sensitivities can exist for other portals, too.

\item {\bf Dark Matter and Invisible Higgs Decays.} In modeling the singlet CFT in (\ref{lagran-singlet}) and constructing the portals in (\ref{portals}), $\Sigma_{\mu}$ has been taken to be a non-gauge real vector field. It is not a gauge field. If it is a gauge field (of some $U(1)$ group in the singlet CFT) then one necessarily puts
$b=0$ and $\lambda_{S1}=0$ in (\ref{lagran-singlet}). Furthermore, it then couples to the $SM$ fields via only the hypercharge channel in the last term of (\ref{portals}). Therefore, letting $\Sigma_{\mu}$ be a non-gauge vector gives a richer class of interactions. This kind of  vectors (and scalars) are known to play decisive roles in developing alternatives to Dark Matter paradigm \cite{bekenstein} as well as in modeling the Dark Matter itself \cite{vec-dm}.

The singlet scalar is particularly interesting. In Standard frame, in general, it interacts as
\begin{eqnarray}
\label{dm-model}
-\sqrt{-g}\left[ \frac{1}{2} g^{\mu\nu} \partial_{\mu} S\, \partial_{\nu} S + \frac{1}{2} \zeta_c R(g) S^2 + \lambda_{H0}  \left(H^{\dagger} H\right) S^2 + \frac{1}{4} \lambda_S S^4 \right]
\end{eqnarray}
in the simpler case of a real scalar field. The conformal transformation of metric in (\ref{conf-trans}) plus the transformation of the scalar field as $S=\left(\frac{\Phi}{M_c}\right) \tilde{S}$, together, transmute (\ref{dm-model})  into the Gravic frame in which $\tilde{S}$ completely decouples from the Higgs field, and acquires the hard mass $\sqrt{\lambda_{0}} M_c$. Dividing $\tilde{S}$ into its long-wavelength component $\tilde{s}$ and short-wavelength component $\delta \tilde{s}$ and integrating out $\delta \tilde{s}$ (having frequencies $\Lambda_{\texttt{EW}}$ up to  $\Lambda_{\texttt{UV}} \sim M_{Pl}$) the effective field theory at the electroweak scale is formed. This effective theory involves renormalized couplings, in particular, the singlet self-couplings
\begin{eqnarray}
\label{renormed-lamphi-GR-dm}
\overline{\lambda_{H0}} &=& \lambda_{H0} + \frac{3 \lambda_S}{(4 \pi)^2}\left[ \frac{{\Lambda_{\texttt{UV}}^2}}{M_{c}^2}  + \lambda_{H0}  \log \frac{\Lambda_{\texttt{EW}}^2}{\Lambda_{\texttt{UV}}^2}\right]\,,\nonumber\\
\overline{\lambda_S} &=& \lambda_S + \frac{9 \lambda_S^2}{(4 \pi)^2} \log \frac{\Lambda_{\texttt{EW}}^2}{\Lambda_{\texttt{UV}}^2}\,.
\end{eqnarray}
In addition, there arises a finite correction
\begin{eqnarray}
\label{renormed-lamphi-GR-quart}
\delta\overline{\lambda_{H}} &=& \frac{1}{(4 \pi)^2}\left[  - \frac{\Lambda_{\texttt{UV}}^4}{M_{c}^4} + 2 \lambda_{H0}
\frac{{\Lambda_{\texttt{UV}}^2}}{M_{c}^2}  + \lambda_{H0}^2  \log \frac{\Lambda_{\texttt{EW}}^2}{\Lambda_{\texttt{UV}}^2}\right]
\end{eqnarray}
to the renormalized Higgs quartic coupling in (\ref{renormed-lamphi-GR}). Having $\Lambda_{\texttt{UV}}$ much smaller than $M_c$, these quantum corrections turn out to be minuscule, and hence, the scalar field theory (\ref{dm-model}) qualifies natural. Having the model stabilized at the electroweak scale, its cosmological and phenomenological predictions become UV insensitive. As is well known \cite{dark}, the scalar field theory (\ref{dm-model}) is a viable model of non-baryonic dark matter. In fact, it has more predictive power than the generic ones \cite{dark} since $S$ has strictly vanishing bare mass. The singlet $\hat{s}$ (defined through (\ref{conf-trans-back})) acquires its mass upon electroweak breaking, $m_{\hat{s}}^2 = \overline{\lambda_0} \langle \phi^2 \rangle$, where the Higgs VEV follows from (\ref{vev-higgs}) with its quartic coupling further corrected by (\ref{renormed-lamphi-GR-quart}). Depending on the masses and couplings, the $SM$ Higgs boson can decay into pairs of $\hat{s}$ whereby enhancing its invisible width. In fact, the model in (\ref{dm-model}) provides a minimal framework for parametrizing the deviation of the LHC Higgs candidate from the $SM$ Higgs boson. In particular, branching fractions of the Higgs boson provide a precise measure of the $SM$ nature of the Higgs boson \cite{invisible,peskin}.

\item {\bf Inflation.}
The inflationary epoch is started by a large vacuum energy density. In this sense, a singlet scalar field as in (\ref{dm-model}), combined
with the vacuum energy $\overline{V_0}$ in $\widehat{SM}$ Standard frame, can give successful inflation. The perfect flatness of the potential 
will be lifted by quantum corrections in (\ref{renormed-lamphi-GR-dm}) and higher-order ones. However, all these happen if $V_0$ is
large enough, and this we know that happens as detailed in Sec. (\ref{sec32}) in discussing ${\mathcal{M}}^2_{\phi}$ \cite{kofman}. Apart from this, inflationary models based on modified gravity theories \cite{modified} are also allowed because for the mechanism to work the gravitational sector does not need be of Einstein-Hilbert form.

\end{itemize}
In this section we have briefly discussed certain overt phenomena. The strictly conformal nature of new physics facilitates models with less free parameters. Nevertheless, the ones touched above as well as several other phenomena need be analyzed in detail in the present framework.

\section{Conclusion}
\label{sec4}
In this work, we have proposed and studied a novel method for naturalizing the $SM$ Higgs
sector. Our approach differs from the existing ones in utilizing gravitational frames instead
of judiciously-introduced new fields and symmetries at the electroweak scale. The essence of the
naturalization mechanism is that a particular conformal transformation with respect to the Higgs field
changes the Higgs interactions with matter fields in such a way that those terms dangerous for naturalness are
restructured to become benign.

In Sec. \ref{sec2},  we have stated the Higgs naturalness problem by explicitly computing the sensitivities of
the model parameters to the UV scale. Following this, in Sec. \ref{sec3}, we have put forth a novel framework
to naturalize the $SM$ Higgs sector. In essence, we have used gravitational frames, frames that are related
through conformal transformations, as a regularization medium. In fact, we have shown that the frame
which decouples Higgs field from matter fields \cite{demir} naturally forbids those terms
dangerous for naturalness. Here, naturalness springs from the scaling symmetry induced by the conformal
transformations. The conformal anomalies break the scaling symmetry softly, that is, by inducing new
interactions of genuinely electroweak size. We have given a detailed discussion of the questions
of unitarity and renormalizability. Then, we have constructed the effective $SM$ theory at the electroweak
scale by determining the effective potential and effective vertices separately.At the end, we have analyzed
$h \rightarrow \gamma \gamma$ decay as a case study.

Again in Sec. \ref{sec3}, concerning vacuum stability, we have shown that the ghosty nature of
the Higgs field in the Gravic frame enforces new physics, if any, to be a CFT. We have 
explored what forms of new physics can add to the $SM$ interactions, and found that sequential
generations, new gauge groups as well as singlet sectors are all admitted. Accordingly, we have
examined a number of phenomena: massive neutrinos, CP violation, Dark Matter and Inflation. In
the present framework, in light of experimental data from LHC and other sources, several phenomena
need be analyzed in detail. 

We conclude that the gravity-enabled method we have proposed can indeed ensure the naturalness of the Higgs
sector with no need to judiciously arranging some new particles or symmetries at the electroweak scale.
The method naturalizes the SM. It also naturalizes extensions of the SM provided that they are CFTs in which all mass
scales are generated by the $SM$ Higgs field. The particle discovered by the LHC \cite{higgs} turns out
to be highly consistent with the $SM$ Higgs boson \cite{consistent}. Therefore, the gravi-naturalization
approach developed here, which demands no new physics for naturalness yet allows for CFT new physics models,
can be a viable model of the electroweak scale. In the present framework, the  Higgs field becomes a
true `God particle' in that it sources all the masses in the particle sector (not excluding particles
beyond the $SM$).

\section{Acknowledgements}
The author is indebted to Levent Akant, Takhmassib Aliev, {\.I}smail Hakk{\i} Duru, Umut G{\"u}rsoy, Ali Kaya, Can Koz{\c c}az, Se{\c c}kin K{\"u}rk{\c c}{\"u}o{\~g}lu, Erkcan {\"O}zcan, Altu{\~g} {\"O}zpineci, Nam{\i}k  Pak, Tongu{\c c} Rador, {\.I}smail Turan, Teoman Turgut and    Kayhan {\"U}lker for  their comments, suggestions and criticisms. He also thanks Asl{\i} Alta{\c s}, Alper Hayreter, Canan Karahan, Beyhan Puli{\c c}e, Selin Soysal and Onur Tosun for discussions on new physics effects.

\end{document}